\documentclass[reqno,12pt]{article}
\usepackage{amsmath,amssymb,amsfonts,epsfig,graphicx,euscript}
\usepackage{graphics}
\newcommand{\bse}{\begin{subequations}}
\newcommand{\ese}{\end{subequations}}
\newcommand{\be}{\begin{equation}}
\newcommand{\ee}{\end{equation}}
\newcommand{\bea}{\begin{eqnarray}}
\newcommand{\eea}{\end{eqnarray}}
\newcommand{\ba}{\begin{array}}
\newcommand{\ea}{\end{array}}

\begin{document}

\begin{flushright}
\end{flushright}
\begin{flushright}
\end{flushright}
\hfill%
\begin{center}

{\LARGE {\sc On Effective Potential in Tortoise Coordinate}}

\bigskip
{M. A. Ganjali\footnote{ganjali@theory.ipm.ac.ir}} \\
{Department of Fundamental Sciences,\\ Tarbiat Moaallem
University,\\P. O. Box 31979-37551, Tehran, Iran}
\\

\end{center}

\bigskip
\begin{center}
{\bf { Abstract}}\\
\end{center}
In this paper, we study the field dynamics in Tortoise coordinate where the equation of motion of a scalar can be written as Schrodinger-like form. We obtain a general form for effective potential by finding the Schrodinger equation for scalar and spinor fields and study its global behavior in some black hole backgrounds in three dimension such as BTZ black holes, new type black holes and black holes with no horizon.

Especially, we study the asymptotic behavior  of potential at infinity, horizons and origin and find that its asymptotic in BTZ and new type solution is completely different from that of vanishing horizon solution. In fact, potential for vanishing horizon goes to a fixed quantity at infinity, while in BTZ and new type black hole we have an infinite barrier.

\newpage
\section{Introduction}
 After finding the first black solution of the general
relativity, an enormous effort has been done for understanding
the various physical properties of such solutions. Some
interesting and extra ordinary results, such as its singularity, its connection to
thermodynamics, ... and recently its holographic nature were found albeit some big puzzles such as black
hole information paradox have not been solved up to now \cite{Maldacena:1997re}.
It is believed that finding the physics laws in its fundamental level
needs the full understanding of black hole physics.

One of such interesting and fundamental properties is that the general relativity is diffeomorphism invariant
and so, frames which are related with such diffeomorphism transformations have similar
physical properties.

One of the important coordinate system is the Tortoise
coordinate. This coordinate system is defined such that one be
able to write down the equation of motion of a scalar field, for
example, like Schrodinger equation
\cite{Harris:2003eg}. Finding and studying physics in this
coordinate system has several advantage. For example, after obtaining that,
one may be able, by solving the Schrodinger equation, to quantize
the system exactly and find the spectrum and Hilbert space of the
theory. Instead of this, because of the effects of the boundary
conditions on the dynamics of the fields, one may interest in the
asymptotic behavior of the field dynamics, \textit{i.e} at far
infinity and near the horizon\footnote{For example, boundary
conditions have important roles in studying the quasi normal
modes of various fields in black hole background. }\cite{Harris:2003eg,Oh:2009if}. In Tortoise
coordinate, this feature can also be addressed by evaluating the
effective potential at boundary of space-time.

Due to the above reasons, we are interested to study some
aspects of the effective potential seen by a Tortoise observer. After a
general discussion on the way for finding the potential in the
next section, we will focus our attention to the three
dimensional black hole solutions and find
the global behavior of the potential for an scalar field. We
will present some examples such as BTZ solution with some details. At
the end, we will demonstrate some comments on the spinor field
potential.

Before starting the next section, a few words about three
dimensional gravity may be useful. It has long been known that
usual three dimensional Einstein-Hilbert gravity has no
propagating degrees of freedom, but the situation is different
when higher curvature terms added to the action\cite{Bergshoeff:2009hq}. Such terms may
be added differently and leads to various gravity theory in three
dimensions. Interestingly, in these new extensions of gravity in
three dimension, one may find massive propagating graviton modes\cite{Bergshoeff:2009hq}. Because
of such properties, Topologically Massive Gravity(TMG), New
Massive Gravity(NMG) and Extended New Massive Gravity\cite{Bergshoeff:2009hq,Moussa:2003fc} are
theories which are interested in recently  and studied extensively
from various point of view such as unitarity, linearization,
supersymmetric extension, black hole solutions, dual conformal
field theory descriptions, new type black holes, quasi normal modes
and \textit{etc}\cite{Oh:2009if,Bergshoeff:2009hq,Moussa:2003fc,Anninos:2008fx}. In this paper, our goal is to study the
potential behaviour seen by a field in three dimensional black
hole background in Tortoise coordinate.

\section{Tortoise Coordinate and Effective Potential}
In this section, we study some general properties of effective
potential seen by a Tortoise observer. First of all, suppose that
the equation of a general field $\Phi(t,r;k_i)=e^{-i\omega
t}\phi(r;k_i)$
in a spherically symmetric space has been written as%
\bea%
\frac{d^2}{dr^2}\phi(r)+A(m,\omega,k_i;r)\frac{d}{dr}\phi(r)+\left(F^2(r)\omega^2+B(m,\omega,k_i;r)\right)\phi(r)=0
\eea%
where $\omega$ is oscillating mode of $\Phi$ and $k_i$ are some
quantum numbers(such as angular momentum). $A(r), B(r)$ and $F(r)$
are some functions appeared in various fields(scalar, spinor, \textit{etc})
equations of motion. Tortoise coordinate is defined by new
variable $r^*$ and new field $\phi^*$ as
\bea%
\phi^*(r^*)=\theta(r)\phi(r),\;\;\;\;\;\;\;\;\;\;r^*=f(r),
\eea%
where the functions $f(r)$ and $\theta(r)$ are uniquely determined
such that one be able to bring the equation of motion of $\phi$ field in
Schrodinger-like form\cite{Harris:2003eg}%
\bea%
(-\frac{d^2}{dr^{*2}}+U(r^*))\phi^*(r^*)=\omega^2\phi^*(r^*)
\eea%
Applying Tortoise change of coordinate and demanding
the coefficient of $\omega$ be equal to one and noting that the
coefficient of $\frac{d}{dr^*}\phi(r^*)$ should be zero, one can
obtain
following conditions on $f(r)$ and $\theta(r)$%
\bea\label{condition1}%
f'&=&\pm F,\\%
\frac{f''}{f'}&=&2\frac{\theta'}{\theta}-A.
\eea%
where the prime denotes derivative respect to $r$. Using these
equations one can find the general form of
effective potential in spherically symmetric space as%
\bea%
U(r)=\frac{1}{2F^2}\left(\frac{F''}{F}-\frac{3}{2}(\frac{F'}{F})^2+A'+\frac{A^2}{2}-2B\right)
\eea%
Using $A(r)$ and $B(r)$ one can obtain the potential for scalar
or other types of fields. Here, we should note that this formula
for effective potential is a function of $r$ instead of $r^*$. So,
for completing the calculation, one should solve the equation
$f'=\pm F$ to find the $r$ dependency of $r^*$ and then finds the
inverse function $r=f^{-1}(r^*)$. This procedure, however, can
not be done analytically for all cases, although there are some
examples in which one can obtain the inverse function exactly
such as BTZ black hole. But, our aim is not finding
exact inverse function $f^{-1}(r^*)$. In fact, we want to analyze
the global behavior of the effective potential, by some simple
calculations such as finding the extremums of potential, without
using the exact coordinate relation $r=f^{-1}(r^*)$. we will do
this in next section with more details.

Note also that, as we can see easily, there is a freedom in
choosing the sign of $f'(r)=\pm F$. In fact, the form of the potential
does not depend on this sign but, if one demands the continuity of
potential functions inside and outside of the horizon at the
horizon one should usually choose different signs for these two
regions. We will come back to this later.
\section{Scalar in 3-D Black Hole Background}%
In this section, we drive the equation of motion of scalar fields in a
general three dimensional spherically symmetric black hole
background. We consider the following form
for the metric
 \bea\label{metric}
ds^2=-N^2(r)dt^2+R^2(r)\left(d\theta^2+N_{\theta}(r)\right)^2+P^2(r)dr^2,
\eea%
The $N(r), R(r), N_{\theta}(r)$ and $R(r)$ are functions of $r$
and are
 specified with the specific solution of equation of motion in three dimensional
 gravity. It is easy to see that $\Delta=\sqrt{-g}=NRP$.\\
The equation of motion of a massive scalar field in curved
background is given by%
\bea \label{eom scalar}
\left(\frac{1}{\sqrt{-g}}\partial_{\mu}(\sqrt{-g}\partial_{\mu}-m^2)\right)\Phi(t,\theta,r)=0,
\eea %
Note that we do not consider the back reaction of the scalar field on the metric.
Using (\ref{metric}), (\ref{eom scalar}) and considering
$\Phi(t,\theta,r)=e^{-i(\omega t-k\theta)}\phi(r)$, one finds%
\bea\label{scalar eom} %
\frac{\partial^2\phi}{\partial
r^2}+\left(\frac{\Delta'}{\Delta}-2\frac{P'}{P}\right)\frac{\partial
\phi}{\partial
r}+\left(\frac{P^2}{N^2}(\omega+kN_{\theta})^2+\frac{P^2}{R^2}(m^2R^2-k^2)\right)\phi=0
\eea %
From this equation one may define effective mode
$\tilde{\omega}=\omega+kN_{\theta}$ and effective mass square
$\tilde{m}^2=m ^2-k^2/R^2$ and for a constant solution one finds
dispersion relation $\tilde{\omega}=\frac{N}{R}\tilde{m}$.

Let us write down the form of the function $U(r)$ for scalar
field using $A$ and $B$ in (\ref{scalar eom})
\bea%
&&A(m,\omega,k;r)=\frac{\Delta'}{\Delta}-2\frac{P'}{P}\cr%
&&B(m,\omega,k;r)=\frac{P^2}{N^2}(2\omega k
N_{\theta}+k^2N^2_{\theta})+\frac{P^2}{R^2}(m^2R^2-k^2).
\eea%
So we obtain%
\bea\label{pot sca}%
U(r)=\frac{N^2}{2P^2}\left(\frac{R''}{R}-\frac{R'}{R}\left(\frac{1}{2}\frac{R'}{R}+\frac{P'}{P}-\frac{N'}{N}\right)-2B\right)
\eea\\%
Using the general equation of motion of components of the metric
field one may rewrite this formula in a more compact form but, we
do not proceed further and only will use the specified form
of $R, P$ and $N$. In the next part of this section we will bring
some examples, BTZ black hole, new type black hole and black hole
with vanishing horizon.

\subsection{BTZ Black Hole}
 The BTZ black hole solution is given by\cite{Banados:1992wn}
 \bea
 &&N^2(r)=\frac{(r^2-r_+^2)(r^2-r_-^2)}{r^2},\;\;\;\;\;\;\;\;\;\;\;\;\;\;R^2(r)=r^2\cr
 &&P^2(r)=\frac{r^2}{(r^2-r_+^2)(r^2-r_-^2)},\;\;\;\;\;\;\;\;\;\;\;\;\;\;N_{\theta}=\frac{r_+r_-}{r^2}
 \eea
 where $r_+$ and $r_-$ are the horizons of the BTZ black hole\footnote{ Throughout of this paper we assume $r_+,r_-\geq 0$}. We
 also choose the $AdS$ length $L=1$.

For studying the effective potential, for simplicity
and getting insight on the problem, we firstly consider the
extremal BTZ black hole although, the results of extremal BTZ can be obtained from non-extremal case
\footnote{In this paper we evaluate the potential for zero mode, $k=0$ but, some results
about global behavior of effective potentials are general.}.

\textit{Extremal BTZ:}
First of all, we calculate $r^*=f(r)$ as%
\bea%
&&r^*=\pm\int{\frac{P}{N}dr}=\pm\int{\frac{r^2}{(r^2-r_0^2)^2}dr}%
\eea%
where $r_+=r_-=r_0$. As we mentioned before, appropriate choosing
of the sign in front of $P/N$ helps us to find $r^*$ in which has
a nice behavior. In fact, by choosing $-$ for $r>r_0$ and $+$ for
$r<r_0$ regions and setting the integration constant to zero
we obtain the following range of changes for $r^*$
\bea%
r=[0,r_0)\;\;\;\;\;\;\;\;\;\;\Leftrightarrow \;\;\;\;\;r^*=[0,+\infty),\cr%
r=(r_0,+\infty)\;\;\;\;\;\Leftrightarrow
\;\;\;\;\;r^*=(+\infty,0].
\eea%
By this notification, we have%
\bea%
r^*=\pm\frac{r}{2(r_0^2-r^2)}\pm\frac{1}{4r_0^2}\ln{\left(\pm\frac{r_0-r}{r_0+r}\right)}
\eea%
where $+$ and $-$ are understood for interior and exterior regions
of black hole respectively. Unfortunately, we can not find the
inverse function $r=f^{-1}(r^*)$. So, we write the effective
potential in term of $r$ and study the general behavior of that.

It is important to note that the asymptotic behavior of $U$ as a function of $r^*$ is similar to that of $U(r)$.
Using (\ref{pot sca}) one may obtain %
\bea%
U(r)=\frac{(r^2-r_0^2)^2}{4r^6}\left((3-4m^2)r^4+2r_0^2r^2-5r_0^4\right).
\eea%
Noting that
$\frac{d}{dr^*}U(r^*)=\frac{1}{f'(r)}\frac{d}{dr}U(r)$, one can
also obtain derivative of potential and set it zero to find extremums of $U(r^*)$.  %
\bea%
\frac{d}{dr^*}U(r^*)=\hspace{9cm}\cr
\frac{(r^2-r_0^2)^2}{2r^9}\left((3-4m^2)r^8+(6+4m^2)r_0^4r^4-24r_0^6r^2+15r_0^8\right)=0
\eea%
As we can see, the asymptotic behavior of $U(r)$ and the extremums of $U(r)$
depend on the range of $m^2$.
\begin{enumerate}
\item{$m^2=0$:}
\bea %
U(r\mapsto 0)\mapsto -\infty,\;\;\;\;%
U(r=r_0)=0,\;\;\;\;%
U(r\mapsto +\infty)\mapsto +\infty.
\eea%
From (15) we see $U(r)$ has only one real positive extremum in
$r=r_0$. In fact, the $r=r_0$ also is an inflection point of
$U(r)$(Figure(1)).
\item{ $0\leq m^2\leq\frac{3}{4}:$}
\bea %
U(r\mapsto 0)\mapsto -\infty,\;\;\;\;%
U(r=r_0,r_2)=0,\;\;\;\;%
U(r\mapsto +\infty)\mapsto +\infty.
\eea%
and potential has two extremums one in $r_1=r_0$ and the other in
$r_2>r_0$(Figure(2)).
\item {$\frac{3}{4}< m^2\leq \frac{3.33}{4}$:}
\bea %
U(r\mapsto 0)\mapsto -\infty,\;\;\;\;%
U(r=r_0,r_2,r_3)=0,\;\;\;\;%
U(r\mapsto +\infty)\mapsto -\infty.
\eea%
and potential has three extremums one in $r_1=r_0$ and two others
are in $r_2,r_3>r_0$(Figure(3)).
\begin{figure}
\centering
\includegraphics{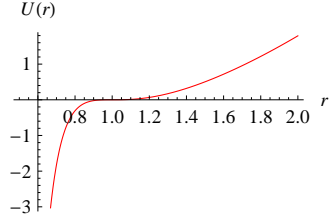}
\caption{Potential for massless scalar, $m^2=0$, in extremal BTZ: $r_0=1$}
\end{figure}
\begin{figure}
\centering
\includegraphics{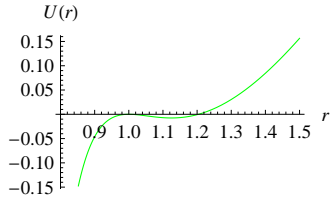}
\caption{Potential for massive scalar $0\leq
m^2=0.5\leq\frac{3}{4}$ in extremal BTZ: $r_0=1$}.
\end{figure}
\begin{figure}
\centering
\includegraphics{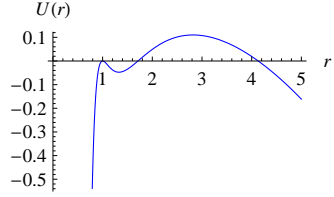}
\caption{Potential for massive scalar $\frac{3}{4}<
m^2=\frac{3.1}{4}\leq \frac{3.33}{4}$ in extremal BTZ: $r_0=1$}
\end{figure}
\item{$\frac{3.33}{4}< m^2$:}
\bea %
U(r\mapsto 0)\mapsto -\infty,\;\;\;\;%
U(r=r_0)=0,\;\;\;\;%
U(r\mapsto +\infty)\mapsto -\infty.\eea%
and potential has only one extremun in $r_1=r_0$(Figure(4)).
\begin{figure}
\centering
\includegraphics{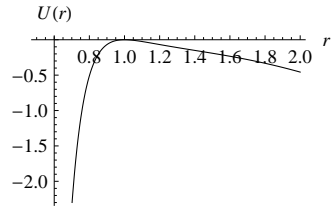}
\caption{Potential for massive scalar $\frac{3.33}{4}< m^2=1$ in
extremal BTZ: $r_0=1$}
\end{figure}
\end{enumerate}

\textit{non-Extremal BTZ:} In this case, by appropriate choose of
the sign of $+$ and $-$
for different regions and solving (\ref{condition1}), one finds the ranges of  $r^*$ as%
\bea%
&&r=[0,r_-)\;\;\;\;\;\;\;\;\;\;\Leftrightarrow \;\;\;\;\;r^*=[0,+\infty),\cr%
&&r=(r_-,r_+)\;\;\;\;\;\;\;\;\Leftrightarrow
\;\;\;\;\;r^*=(+\infty,-\infty),\cr%
&&r=r_+,\infty)\;\;\;\;\;\;\;\;\;\Leftrightarrow
\;\;\;\;\;r^*=(-\infty,0],
\eea%
The effective potential also reads as %
\bea%
U(r)=\frac{(r^2-r_+^2)(r^2-r_-^2)}{r^6}\left((3-4m^2)r^4+(r_+^2+r_-^2)r^2-5r_+^2r_-^2\right).
\eea%
Let us consider only the massless case.
It is obvious that $U(r)$, in massless case, has three real positive zeros at
$r_{\pm}$ and
$$r_0=-\frac{1}{6}(r_+^2+r_-^2-\sqrt{r_+^4+r_-^4+62r_+^2r_-^2}).$$
It is also easy to show that $r_-<r_0<r_+$. We observe that
the asymptotic behavior of $U$ is such that $U\rightarrow -\infty
$ as $r\rightarrow 0$ and $U\rightarrow +\infty $ as $r\rightarrow
+\infty$. Then, we obtain the
extremums of  $U(r^*)$ by solving%
\bea%
\frac{d}{dr^*}U(r^*)=\frac{(r^2-r_+^2)(r^2-r_-^2)}{2r^9}\times\hspace{6cm}\cr%
\times\left(3r^8+(r_+^4+r_-^4+4r_+^2r_-^2)r^4-12r_+^2r_-^2(r_+^2+r_-^2)r^2+15r_+^4r_-^4\right)=0.
\eea%
Again $r_+$ and $r_-$ are extremums of $U(r^*)$. Defining $x=r^2$
and considering the general properties of quartic equations, one
can prove that there are exactly two other real positive extremas
for effective potential. In fact, if $x_1,x_2,x_3$ and $x_4$ be
the
zeros of %
\bea%
3x^4+(r_+^4+r_-^4+4r_+^2r_-^2)x^2-12r_+^2r_-^2(r_+^2+r_-^2)x+15r_+^4r_-^4=0.
\eea%
they should satisfy %
\bea\label{quartic BTZ}%
&&x_1x_2x_3x_4=15r_+^4r_-^4>0\cr %
&&\Sigma x_ix_jx_k=12r_+^2r_-^2(r_+^2+r_-^2)>0\cr%
&&\Sigma x_ix_j=r_+^4+r_-^4+4r_+^2r_-^2>0\cr%
&&\Sigma x_i=0
\eea%
For proving the statement, one should recall that if a complex
number $z$ be a solution of a polynomial equation then the
$\bar{z}$ also is a solution. After all, we can plot the $U(r)$
qualitatively as(Figure(5)).
\begin{figure}
\centering
\includegraphics{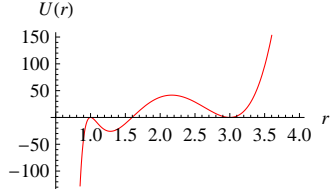}
\caption{Potential for massless scalar in non-extremal BTZ: $r_-=1,r_+=3$.}
\end{figure}
\subsection{New Type Black Hole}
New Type black hole solution is\cite{Bergshoeff:2009hq}
given by\newpage
 \bea
 &&N^2(r)=r^2+br+c,\;\;\;\;\;\;\;\;\;\;\;R^2(r)=r^2\cr
 &&P^2(r)=\frac{1}{r^2+br+c}\;\;\;\;\;\;\;\;\;\;\;\;N_{\theta}=0
 \eea
For non-extremal new type black hole($b^2-4ac\neq 0$), one can find %
\bea%
r^*=\pm\frac{1}{\sqrt{b^2-4ac}}\ln{\left(\frac{2ar+b-\sqrt{b2-4ac}}{2ar+b+\sqrt{b2-4ac}}\right)}
\eea%
and %
\bea\label{pot new}%
U(r)=\frac{1}{2r^2}(ar^2+br+c)\left((3a-2m^2)r^2+br-c\right)
\eea%
Let us consider the extremal case where $b^2-4ac=0$ and $r_+=r_-=r_0=-\frac{b}{2a}$. In this case we can find the inverse function $r=f^{-1}(r^*)$ exactly. In fact, we have
\bea%
r^*=\pm\int{\frac{1}{(r-r_0)^2}}=\mp\frac{1}{r-r_0}+c_0,
\eea%
where $c_0$ is an integration constant and $+/-$, in front of the integral, are for interior and exterior regions of horizon respectively. Here, we set $c_0=\frac{1}{r_0}(1-\theta(r-r_0))$ where $\theta(r-r_0)$ is \textit{step} function\footnote{Note that even with such $c_0$ constant, $r^*(r_0)=+\infty$ and the potential is continious.}. So, we obtain the domain of changes of $r^*$ as
\bea%
r=[0,r_0)\;\;\;\;\;\;\;\;\;\;\Leftrightarrow \;\;\;\;\;r^*=[0,+\infty),\cr%
r=(r_0,+\infty)\;\;\;\;\;\Leftrightarrow
\;\;\;\;\;r^*=(+\infty,0].
\eea%
Also, we can easily find
\bea\label{invers}%
r=\mp\frac{1}{r^*-c_0}+r_0.
\eea%
Putting (\ref{invers}) in (\ref{pot new}), one obtains $U(r^*)$. Let us study the asymptotic of $U$, again, in $r$ coordinate. We have different regions due to mass parameter as
\begin{enumerate}
\item{$3a-2m^2\leq 0$:}
\bea %
U(r\mapsto 0)\mapsto -\infty,\;\;\;\;%
U(r=r_0)=0,\;\;\;\;%
U(r\mapsto +\infty)\mapsto -\infty.
\eea%
$U(r)$ has only one real positive extremum in $r=r_0$(Figure(6)).
\begin{figure}
\centering
\includegraphics{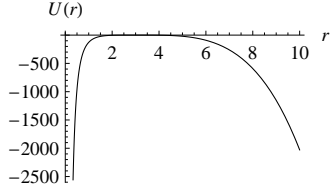}
\caption{Potential for massive scalar in extremal new type B.H:
$r_0=1,3a-2m^2=-1$}
\end{figure}
\item{ $0< 3a-2m^2\leq 3:$}
\bea %
U(r\mapsto 0)\mapsto -\infty,\;\;\;\;%
U(r=r_0)=0,\;\;\;\;%
U(r\mapsto +\infty)\mapsto +\infty.\eea%
and potential has two extremums one in $r_1=r_0$ and the other
in $r_2>r_0$(Figure(7)).
\begin{figure}
\centering
\includegraphics{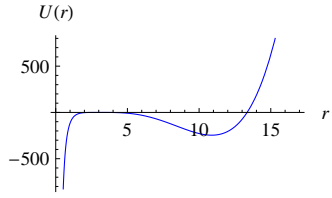}
\caption{Potential for massive scalar in extremal new type B.H:
$r_0=1,3a-2m^2=0.5$}
\end{figure}
\item {$3< 3a-2m^2$:}
\bea %
U(r\mapsto 0)\mapsto -\infty,\;\;\;\;%
U(r=r_0)=0,\;\;\;\;%
U(r\mapsto +\infty)\mapsto +\infty.\eea%
and potential has two extremums one in $r_1=r_0$ and the other is
in $r_2<r_0$(Figure(8)).
\begin{figure}
\centering
\includegraphics{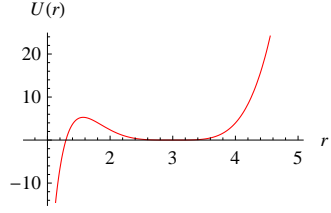}
\caption{Potential for massive scalar in extremal new type B.H:
$r_0=1,3a-2m^2=10$ in extremal BTZ: $r_0=1$}
\end{figure}
\end{enumerate}

\subsection{Black Hole with Vanishing Horizon} In this section,
we study black holes with vanishing horizon in three dimensions.
We consider following case where the solution has a null Killing
vector\cite{Clement:2009ka}. As we will see, the global behavior of this solution, in
general, is different from the BTZ and new type solutions. In
fact, the effective potential, with a certain condition, has two
extremum(one in origin and the other out of origin), otherwise,
has only one extremum at origin and so is a monotonic function. A class of
such solutions in three dimensional massive gravity is given by\cite{Clement:2009ka} %
 \bea
 &&R^2(r)=l^2a_+r^2+2r+l^2d,\;\;\;\;\;\;\;\;\;\;\;N^2=\frac{4r^2}{l^2R^{2}}\cr
 &&P^2(r)=\frac{l^2}{4r^2}\;\;\;\;\;\;\;\;\;\;\;\;\;\;\;N_{\theta}=\frac{2r}{lR^2}
 \eea
where $l,m$ are parameters in three dimensional massive gravity
and are related to cosmological constant and mass parameter of the action and $a_+, d$ are some integration constants. Having regular
solution forces us $m^2l^2=17/2$ and $a_+>0, d>0$. Defining
$l^2a_+r=x$ and $l^4da_+=s$ one obtains for $r^*$ as%
\bea\label{vanish r}%
r^*=+\frac{l^5a_+^{\frac{3}{2}}}{4}\int{\frac{\sqrt{x^2+2x+s}}{x^2}dx}
\eea%
Here one can find the exact result of the integral (\ref{vanish r}). By the above choosing of $+$ in front of the integral we will have
\bea%
r=[0,+\infty)\;\;\;\;\;\;\;\;\;\;\Leftrightarrow \;\;\;\;\;r^*=(-\infty,+\infty).%
\eea%
Then, we obtain the potential for massless scalar as
\bea%
U(r)=\frac{4x^3}{a_+l^6(x^2+2x+s)^3}\left(x^3+6x^2+(6s+3)x+4s\right)
 \eea%
It is easy to see that $U(r)\geq 0$ for $r\geq 0$ and %
\bea%
U(r\rightarrow 0)\rightarrow 0,\;\;\;\;\;\;\;\;U(r\rightarrow
+\infty)\rightarrow \frac{4}{a_+l^6}
\eea%
Also, for derivative of potential $\frac{d}{dr^*}U(r^*)$, one can
obtain%
\bea%
\frac{d}{dr^*}U(r^*)\varpropto
\frac{x^4}{R^9}\left((3-s)x^3+(5s+1)x^2+2s(2s+1)x+2s^2\right)
\eea%
As we can see $x=0$ and roots of%
\bea\label{cubic non-v}%
(3-s)x^3+(5s+1)x^2+2s(2s+1)x+2s^2=0
\eea%
are the extremum of $U$.
Here there are two regions in term of
$s$.
\begin{itemize}
\item $0\leq s\leq 3$: In this case, one can easily see that equation
(\ref{cubic non-v}) has no real positive solution because all terms are positive. So the general behavior of $U$ can be plotted as blue curve in(Figure(9)).
\item $3<s$: Here, by considering the
general properties of cubic equations
\bea\label{qubic}%
&&x_1x_2x_3=-\frac{2s^2}{3-s}>0\cr %
&&\Sigma x_ix_j=\frac{2s(2s+1)}{3-s}<0\cr%
&&\Sigma x_i=-\frac{5s+1}{3-s}>0
\eea%
 one can prove that
(\ref{cubic non-v}) has only one real positive solution. So, one
may plot the effective potential as red curve in(Figure(9)).
\begin{figure}
\centering
\includegraphics{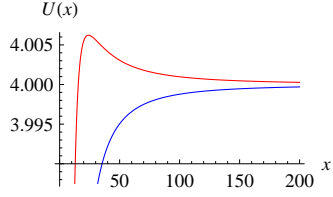}
\caption{Potential for massless scalar in black hole with no
horizon: the blue curve is for $0<s\leq3$ and the red curve is
for $s>3$.}
\end{figure}
\end{itemize}
\section{Fermions in $3-$D Black Hole Background}
In this section, we study the potential seen by a Tortoise observer for a spinor field. Spinor field representation in three dimension has two components and the coupled equation of motion for these fields is given by Dirac equation in
curved background as\cite{Harris:2003eg}%
\bea%
(-i\gamma^ae_a^{\mu}D_{\mu}+m)\Psi(t,\theta,r)=0,
\eea%
where the covariant derivative is
$D_{\mu}=\partial_{\mu}-\frac{i}{4}\eta_{ac}\omega^{c}_{b\mu}\sigma^{ab}$
and $\sigma^{ab}=\frac{i}{2}[\gamma^a,\gamma^b]$.
$\omega^{c}_{b\mu}$ are spin connection and are defined by
veilbeins and Christoffel coefficients as
$\omega^{c}_{b\mu}=e^c_{\nu}\partial_{\mu}e^{\nu}_b+e^c_{\nu}e^{\sigma}_b\Gamma^{\nu}_{\sigma\mu}$.
For the background (\ref{metric}), one can obtain non-zero veilbains as%
\bea%
e^t_0=-\frac{1}{N},\;\;\;\;\;e^{\theta}_0=\frac{N_{\theta}}{N},\;\;\;\;\;e^r_1=\frac{1}{P},\;\;\;\;\;e^{\theta}_2=\frac{1}{R},
\eea%
and so the covariant derivatives take the following forms %
\bea%
&&D_t=\partial_t,\;\;\;\;\;\;D_{\theta}=\partial_{\theta},\cr%
 &&D_r=\partial_r-\frac{i}{8NR}\partial_r(R^2N_{\theta})\sigma_1
\eea%
By rewriting two dimensional spinor field $\Psi(t,\theta,r)$
as%
\bea%
\Psi(t,\theta,r)=e^{-i(\omega t-k\theta)}
\begin{pmatrix}
  \phi^+ \\
  \phi^-
\end{pmatrix}
\eea%
one finds %
\bea%
&&{\cal{M}}\phi^++{\cal{N}}\phi^-+\partial_r\phi^-=0\cr%
&&\bar{\cal{M}}\phi^-+\bar{\cal{N}}\phi^++\partial_r\phi^+=0%
\eea%
where%
\bea%
&&{\cal{N}}=i\frac{P}{N}(\omega+kN_{\theta})\cr%
&&{\cal{M}}=\frac{1}{8NR}\partial_r(R^2N_{\theta})+P(m+i\frac{k}{R}),%
\eea%
and one may obtain following tow decoupled equations for $\phi^+$
and
$\phi^-$%
\bea%
&&\partial^2_r\phi^++(-\frac{\partial_r{\bar{\cal{M}}}}{\bar{\cal{M}}})\partial_r\phi^++(|{\cal{N}}|^{2}-|{\cal{M}}|^{2}+
\bar{\cal{M}}\partial_r(\frac{\bar{\cal{N}}}{\bar{\cal{M}}}))\phi^+=0,\\
&&\partial^2_r\phi^-+(-\frac{\partial_r{{\cal{M}}}}{{\cal{M}}})\partial_r\phi^-+(|{\cal{N}}|^{2}-|{\cal{M}}|^{2}+
{\cal{M}}\partial_r(\frac{{\cal{N}}}{{\cal{M}}}))\phi^-=0%
\eea%
Note that the coefficient of $\omega^2$ again is
$\frac{P^2}{N^2}$. This is just from the fact that spinor equation
is a linearization of Schrodinger equation. So, for evaluating the
effective potential of spinors we have for $\phi^+$
\bea%
&&A^+(m,\omega,k;r)=-\frac{\partial_r{\bar{\cal{M}}}}{\bar{\cal{M}}}\cr%
&&B^+(m,\omega,k;r)=\frac{P^2}{N^2}(2\omega k
N_{\theta}+k^2N^2_{\theta})-|{\cal{M}}|^{2}+
\bar{\cal{M}}\partial_r(\frac{\bar{\cal{N}}}{\bar{\cal{M}}})
\eea%
and simply $A^-=\bar{A}^+, B^-=\bar{B}^+$ for $\phi^-$. Thus, we
obtain
\bea%
U(r)=\hspace{19cm}\cr%
\frac{N^2}{2P^2}\left(\frac{P''}{P}-\frac{N''}{N}-\frac{\bar{\cal{M''}}}{\bar{\cal{M}}}-\frac{3}{2}(\frac{P'}{P})^2+\frac{1}{2}(\frac{N'}{N})^2
+\frac{3}{2}(\frac{\bar{\cal{M'}}}{\bar{\cal{M}}})^2+(\frac{P'}{P})(\frac{N'}{N})-2B\right)(21)\hspace{6cm}
\eea%
for $\phi^+$ and $\bar{U}$ for $\phi^-$. Here, we do not want to
study the potential with details and only present some comments for two cases.
\begin{itemize}
\item BTZ and New type Black Holes: We see that for BTZ and new type black holes $\partial_r(R^2N_{\theta})=0$. Then, massless spinors components
$\phi^+$ and $\phi^-$ have the same potential but only the zero of them have decoupled dynamics from each other. Beside of these, for massless zero mode we obtain a harmonic oscillator-like potential for two components.

In massive case, two components always have coupled dynamics and only stationary zero modes they have the same potential.

\item Black Holes with No Horizon: But, in black hole with no horizon where we have $\partial_r(R^2N_{\theta})=2/l\neq 0$, even the stationary massless zero modes do not have decoupled dynamics.
    They also do not have the same potential due to the fact that ${\cal{N}}$ is imaginary unless for stationary zero mode where.
    Especially for stationary massless zero mode they have an oscillator-like potential.
%
%
%
\end{itemize}
\section{Conclusion}
In this note, we study the fields dynamics in Tortoise frame. In this frame, by defining new coordinate $r^*(r)$ and an unknown auxiliary field $\theta(r)$ the equation of motion of a scalar field can be written as Schrodinger-like form. Using this change of variables we can obtain two conditions for $r^*$ and $\theta$ and then obtain a general form for effective potential. Then, we apply this formula for scalar and fermionic fields in some black hole backgrounds in three dimension such as BTZ black holes, new type black holes and black holes with no horizon and find the potential. Although, in general, we can not find the inverse function $r(r^*)$ but by some simple calculations, We study the asymptotic behavior of potential at infinity, horizons and origin and also find the extremum of that.

Interestingly, we find that the asymptotic of potential in BTZ and new type solution is completely different from that of vanishing horizon solution. In fact, potential for vanishing horizon goes to a fixed number at infinity but for BTZ and new type black hole background we have an infinite barrier at infinity.

\section*{Acknowledgment}
I would like to thank Dr Fareghbal for useful comments.
I also like to thank Dr Kanjouri due to his guides in plotting the figures.

This work is supported by the "Grant for Research Projects" in Tarbiat Moallem University.

\end{document}